# PHOTOACTIVE GOLD NANOPARTICLE SOFTOXOMETALATES (SOM) USING A KEPLERATE FOR SYNTHESIS OF POLYSTYRENE LATEX MICROSPHERES BY PHOTO-POLYMERIZATION

ATHARVA SAHASRABUDHE

*Eco-friendly Applied Materials Laboratory (EFAML), Materials Science Centre, Department of Chemical Sciences, Indian Institute of Science Education and Research-Kolkata, Mohanpur Campus*
*PO: Krisi Viswavidyalaya,*
*Mohanpur - 741252,*
*Nadia, West Bengal, India*

SOUMYAJIT ROY\*

*Eco-friendly Applied Materials Laboratory (EFAML), Materials Science Centre, Department of Chemical Sciences, Indian Institute of Science Education and Research-Kolkata, Mohanpur Campus*
*PO: Bidhan Chandra Krisi Viswavidyalaya,*
*Mohanpur - 741252,*
*Nadia, West Bengal, India,*
*\*Email: s.roy@iiserkol.ac.in.*

A green and facile synthetic protocol for the preparation of photoactive gold nanoparticle-soft oxometalates (AuNP-SOM) using a unique Keplarate type oxomolybdate cluster viz. $\{Mo_{132}\}$, is reported. The as synthesized AuNP-SOMs are fully characterized by Electronic absorption spectroscopy, Raman spectroscopy and High resolution transmission electron microscopy (HR-TEM). We further demonstrate that by merely tuning the ratio of the precursors in solution it is possible to control the morphology of AuNP-SOM nanostructures. Moreover, these photoactive AuNP-SOMs are employed as photocatalysts for the photopolymerization of styrene to generate colloidal monodisperse polystyrene microspheres in a controllable way in the absence of any external co-initiator or inert atmosphere. Effect of styrene concentration on the size of microspheres is studied using DLS and optical microscopy. Finally, the role of AuNP-SOMs as efficient photocatalyst is established through several control experiments and a reaction pathway is proposed.

*Keywords*: Soft-oxometalates, $\{Mo_{132}\}$, Keplerates, Photocatalysis, Photopolymerization, Polystyrene Latexes.

## 1. Introduction

Latex microspheres are ubiquitous and have a wide range of applications starting from targeted drug delivery, coating, adhesive, biotechnique, catalyst supports and allied systems.[1-11] They are cheap and are usually made by a diverse range of methodologies, like emulsion polymerization, seeded emulsion polymerization, emulsifier free copolymerization, precipitation polymerization, dispersion polymerization and so on.[12-27] However, owing to their ever increasing demand and drive for greener and cheaper printable electronics and allied applications, there is an increasing demand for greener and faster synthetic methodologies for latexes. Questions arise: is it possible to synthesize Latex Microspheres using a fast, facile and green 'zero-VOC'-synthetic route? Can such latex microspheres be prepared controllably under one-pot condition using light, in the absence of any co-initiator and under open atmosphere conditions? Here we report such a synthesis of polystyrene latex microspheres by photo-polymerization assisted by photoactive softoxometalates (SOMs).

SOMs have been defined as a class of materials which are colloidal hence soft super-structures of oxometalates that transcend the translational periodicity limit of polyoxometalates and hence can have aperiodic structures.[28,29] SOMs exist in dispersion and can be spontaneously formed or their structure can be directed. Here we take the second route i.e., directed route for preparing SOMs, which are used as photocatalysts for the synthesis of polystyrene latexes. The synthesis of SOM is carried out with an eye that it caters the following criteria: 1) It is soluble/dispersible in a green solvent like water; 2) The SOM is photo-active; 3) It can initiate photo-polymerization on a fast time-scale. We now explain the choice of precursors for the synthesis of such photoactive, designed SOMs. Au-nanoparticles (Au-NPs) are chosen as a starting precursor. The reason for this choice is also manifold. We explain this choice here briefly. Au-NPs are known to have unique electronic and opto-catalytic effects due to quantum confinements. Hence these properties are highly sensitive to size, shape and morphology of the nanostructures.[30-32] Consequently, to synthesize latex microspheres photo-chemically, it is important to synthesize well-defined precursors with well-defined electronic and opto-catalytic properties. Consequently, Au-NPs with desired sizes and shapes were chosen as a precursor. However, most of the synthetic methods for Au-NPs make use of an organic environment, multiple reagents and a relatively high temperature, thus making them complex and environmentally unfriendly protocols. Hence, to circumvent this problem we synthesized Au-NP based SOM in water using a green synthetic protocol.[33] Such Au-NP based SOMs enable exploration for the synthesis of Au-NPs with a full morphological control by employing mild, environmentally-friendly conditions in a green solvent: water, and also at room temperature.[34] Furthermore the method enables the use of a nontoxic Keplerate $\{Mo_{132}\}$-type giant oxo-molybdate molecular cluster as reducing and stabilizing agents. We now explain the choice of $\{Mo_{132}\}$ Keplerate type cluster as reducing and stabilizing agent.

$\{Mo_{132}\}$ Keplerate type cluster is an unusual anion of the size comparable to that of a protein![35] The ion is unusually large and also possesses 60 $Mo^V$ centres, which can in principle donate 60 electrons to the metal ions and consequently get oxidized. Moreover, $\{Mo_{132}\}$ anion can also act as a bifunctional catalyst by reducing gold cations and then strongly adsorbing on the metal surface; thereby stabilizing the nanostructures via electrostatic repulsions. All these properties render it as an attractive precursor for the design of SOMs. However, our overall design strategy is a combination of existing routes for photochemical generation of AuNPs using oxometalates and use of oxometalates in stabilization of AuNPs.[36-44] In this work we combine these two approaches and extend it for the synthesis of polystyrene latex microspheres. Now we take a look at how the reactions take place and how the synthesis of polystyrene latex microsphere is achieved.

## 2. Experimental

### 2.1. Synthesis of AuNP-SOMs

The oxometalate $\{Mo_{132}\}$ Keplerate, [≡(NH$_4$)$_{42}\{Mo^{VI}_{72}Mo^V_{60}O_{372}(CH_3COO)_{30}(H_2O)_{72}\}$. ca. 250H$_2$O] was synthesized according to the literature procedure.[35] Briefly, after adding N$_2$H$_6$SO$_4$ (6.1 mmol) to a solution of (NH$_4$)$_6$ Mo$_7$O$_{24}$ .4H$_2$O (4.5 mmol) and CH$_3$COONH$_4$ (162.2 mmol) in H$_2$O (250 mL) and stirring for 10 min (color change to blue-green), 50% (v/v) CH$_3$COOH (83 mL) is added. The reaction solution, now green, is stored in an open 500-mL Erlenmeyer flask at 200 °C without further stirring (slow color change to dark brown). After 4 days the precipitated red-brown crystals are filtered off through a glass frit (D2), washed with 90% ethanol and diethyl ether, and finally dried in air.

3 ml of 0.035 mM aqueous solution of $\{Mo_{132}\}$ was added to 4ml of 0.58 mM of aqueous HAuCl$_4$ solution under rapid stirring. The solution was stirred for further 15 minutes. A change in the solution color from brown to pink indicated the formation of AuNP-SOMs. The as prepared AuNP-SOMs were purified by multiple centrifugation-resuspension cycles using ethanol and distilled water. The purified AuNP-SOMs were finally dispersed in 5ml of distilled water. The absence of precipitate for more than 3 months in the AuNP-SOM dispersion is noteworthy.

### 2.2. Synthesis of Citrate stabilized gold nanoparticles

Citrate stabilized gold nanoparticles were synthesized using the Turkevich method.[45] Briefly, 1 ml of 0.5 mM HAuCl$_4$ aqueous solution was added to 18 ml of distilled H$_2$O under rapid stirring and was heated until it began to boil. 1 ml of 0.5 % sodium citrate solution was then added as soon as boiling commenced. Heating was continued until the color changed to pale purple, upon which heating was stopped. Stirring was continued until the reaction mixture cooled down to room temperature. The as prepared nanoparticles were purified by multiple centrifugation-resuspension cycles. The purified nanoparticles were finally dispersed in 10 ml of distilled water.

### 2.3. Characterization of AuNP-SOMs

#### 2.3.1. UV-Vis spectroscopy

The formation of AuNP-SOMs was probed extensively using UV-Vis spectroscopy. Both, time dependent and concentration dependent experiments were performed to understand the formation of the AuNP-SOMs. For the time-dependent experiments, 0.035 mM aqueous solution of $\{Mo_{132}\}$ was added to 0.58 mM aqueous HAuCl$_4$ solution under rapid stirring. Aliquots were taken from this solution at different time intervals and their UV-Vis spectra were recorded. For concentration dependent experiments, we defined an excess parameter $\Psi = [HAuCl_4] / [\{Mo_{132}\}]$. Typically, the $\Psi$ values were varied from 20 to 35 and corresponding UV-Visible spectra were recorded. All the UV-visible spectra were taken at 25 °C with a Perkin-Elmer Lambda-20 spectrometer.

### 2.3.2. *Raman Spectroscopy*

Time dependent Raman spectra were recorded to monitor the oxidation of $\{Mo_{132}\}$ in solution as the quasi-spherical (icosahedral) construction of these clusters provides an easy Raman spectroscopic 'finger-print' for studying their stability. Typically, 0.035 mM aq.solution of $\{Mo_{132}\}$ was added to 0.5 mM aqueous $HAuCl_4$ solution under rapid stirring. Aliquots were taken from this solution at different time intervals and their Raman spectra were recorded.

### 2.3.3. *Transmission Electron Microscopy*

Transmission electron microscopy (TEM) observations were performed with a JEOL 100CXII transmission electron microscope operating at an accelerating voltage of 100 kV. The sample drops were deposited and dried on a carbon-coated copper grid.

### 2.4. *Photo-polymerization using AuNP-SOMs*

1ml of the as-prepared pure AuNP-SOM dispersion was added to 8 ml of DMSO in a quartz tube. To this solution was added 1ml of styrene, and the mixture was irradiated in a Luzchem Photoreactor (80 W, UVA lamp) for 1 hr. The initial clear solution gradually turned turbid and finally, after 1hr irradiation, the solution turned milky-white, indicating the formation of polystyrene latexes. On completion of the reaction, the reaction mixture is further allowed to stand for 45-60 minutes, during which AuNP-SOM photocatalysts aggregates and settles at the bottom of the reaction tube. It is then separated by simple decantation of the reaction mixture.

### 2.5. *Characterization of Polymer Latex*

### 2.5.1. *Dynamic Light Scattering*

To understand the effect of monomer (styrene) on the size of polystyrene microspheres, concentration-dependent Dynamic Light Scattering experiments were performed using a Malvern Zetasizer Nano. Typically, 1ml of the as-prepared pure AuNP-SOM dispersion was added to 8 ml of DMSO in a quartz tube. To this solution was added varying amounts of styrene, viz. 200μl, 500μl, 700μl and 1000μl and the mixture was irradiated in a Photoreactor (80 W, UVA lamp). Aliquots were taken out at regular time intervals and DLS measurements were performed.

### 2.5.2. *Infrared Spectroscopy*

FT-IR spectrum of the polystyrene microspheres was recorded under attenuated total reflectance mode (ATR-FTIR) with a Nicolet 380 FT-IR spectrometer in the range of 4000–500 cm$^{-1}$. For sample preparation, 1ml of the latex dispersion was dissolved in 3ml chloroform and the solution was transferred onto a separate IR reflective glass slide for measurements.

### 2.5.3. *Optical Microscopy*

Optical Microscope Images of as prepared polystyrene latexes were obtained with a Nikon ECLIPSE 90i microscope with a motorized focusing stage. For sample preparation, a drop of latex dispersion was placed on a glass slide and covered with a cover-slip. The as prepared slide was then mounted onto the microscope for imaging.

### 2.5.4. *Transmission Electron Microscopy*

Polystyrene microspheres were diluted in ethanol and drops were deposited and dried on a carbon-coated copper grid and imaged using a JEOL 100CXII transmission electron microscope operating at an accelerating voltage of 100 kV.

### 3. **Results and Discussion**

AuNP-SOMs are prepared by a facile one-step reduction of $HAuCl_4$ in an aqueous solution with $\{Mo_{132}\}$ Keplerate type cluster acting as a reducing and stabilizing agent. This reaction is fast and it leads to immediate formation of purple AuNPs, which are characterized by a host of techniques, from light scattering to microscopy. The oxidized oxometalate fragments in turn adsorb onto the AuNP surface thereby stabilizing the AuNP-SOMs via electrostatic interactions. This dispersion is then loaded with styrene and exposed to UV light. This in turn leads to almost immediate milky white colouration due to the formation of polystyrene (PS) latex microspheres, which have also been characterized using a host of techniques. This synthetic methodology enjoys certain unique advantages. For instance, as i) the whole preparation process is in water, at room temperature, under atmospheric conditions, thus adhering fully to "green chemistry type" conditions; ii) since the oxometalate contains multiple Mo$^V$ centers which can reduce the metal (Au$^{3+}$) cations, the synthesis is one-pot without any pre-treatment of oxometalate via electrochemical or photochemical methods. ii) The presence of 60 Mo$^V$ metal centers in {Mo$_{132}$}, which can undergo single-electron transfer reactions to reduce the Au$^{3+}$ cations, renders the reaction kinetics of gold nanostructures very fast iii) {Mo$_{132}$} anion plays the role of a 'Janus' catalyst by reducing gold cations and then strongly adsorbing on the metal surface, acting as a oxometalla-surfactant, thereby stabilizing the nanostructures via electrostatic repulsions. iv) The morphology of the so formed nanostructures can be controlled by selecting an appropriate concentration of {Mo$_{132}$} and/or Au$^{3+}$ ions. We now describe the synthetic procedures and obtained results in more details.

### 3.1. *On synthesis of AuNP-SOMs starting from {Mo$_{132}$} Keplerate*

The synthesis of (colloidal) AuNP-SOMs was carried out by simple mixing of aqueous {Mo$_{132}$} and HAuCl$_4$ solutions, followed by stirring for 2 to 3 minutes at room temperature[33], in air, indicating fast reaction kinetics following the equation:

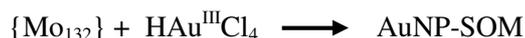

{Mo$_{132}$} + HAu$^{III}$Cl$_4$ ⟶ AuNP-SOM

The reaction has been schematically shown in Fig 1. The formation of a stable AuNP-SOM dispersion can be easily detected by a distinct color change when the initial reddish-brown color of the solution changes to pink or violet (inset Fig 1). This is also confirmed by the scattering of the laser beam passing through the dispersion after the reaction.

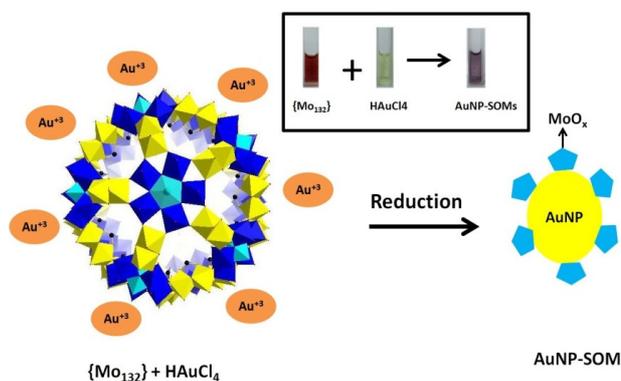

Fig. 1. Formation of AuNP-SOM using Keplarate {Mo$_{132}$} as reducing and stabilizing agent (blue octahedra :Mo$^{VI}$O$_6$ units; yellow octahedral:{ Mo$^V_2$O$_4$} linker units; green pentagonal bipyramids: central MoO$_7$ units). Inset: color change from brown to violet indicating formation of AuNP-SOM.

### 3.2. *On Monitoring the synthesis of AuNP-SOMs starting from {Mo$_{132}$} Keplerate*

With the end in view to monitor the formation of AuNP-SOMs we undertook extensive electronic absorption spectroscopic measurements as HAuCl$_4$ is reduced and stabilized with Keplerate {Mo$_{132}$}. This is because AuNPs are known to have surface Plasmon resonance bands around 530 nm[46-47] while the {Mo$_{132}$} Keplerate has an IVCT band around 450 nm.[35] We first undertook time dependent UV-Visible spectroscopy experiments in order to probe the oxidation of {Mo$_{132}$} and the corresponding formation of AuNP-SOMs. It is observed that with time the IVCT band of the Keplerate at 450 nm decreases in intensity and finally disappears after 4-5 minutes while a new band starts appearing at 542 nm, whose intensity keeps on increasing with time (Fig. 2). The gradual decrease in the intensity and disappearance of the IVCT peak of Mo$_{132}$ clearly indicates the complete oxidation of the Mo$^V$ linkers. The new broad peak appearing at 542 nm is due to the Surface Plasmon Resonance of the so formed AuNP-SOMs and is indicative of a typical dipole resonance associated with spherical or quasi spherical AuNPs.[46-47] It also implies that the reaction for the formation of AuNP-SOMs from the Keplerate proceeds in stoichiometric proportions. This in turn implies that in AuNP-SOM, the Keplerate cluster structure is slowly lost with time. This is also confirmed from time dependent Raman spectroscopy (Fig. 3), where the Raman signature of the skeletal framework of {Mo$_{132}$} is lost upon synthesis of AuNP-SOMs, indicating that probably smaller fragments of oxomolybdates are present on the AuNP-SOM surface. The {Mo$_{132}$} type clusters' decomposition is detected by the change in Raman spectrum induced by the change of the normal coordinates of [O$_{372}$] framework, which contribute to the breathing vibration of the oxomolybdate skeleton. The Raman spectrum of the {*Mo$_{132}$*} cluster shows four bands[35], which have been assigned as following: 950 $cm^{-1}$ [$v(Mo = O_t)$], 374 $cm^{-1}$ [$\delta(Mo = O_t)$], 314 $cm^{-1}$ [$\delta(Mo - O - Mo)$] and the most intense band at 880 $cm^{-1}$ [$v(O_{bri} breathing)$] that corresponds to the totally symmetric $A_{1g}$ breathing vibration of the capsule. The reduction in the intensity of the four major Raman bands clearly indicates degradation of the cluster framework and this is attributed to the oxidation of the Mo$^V$ linkers of the clusters which are used up in AuNP-SOM formation process.

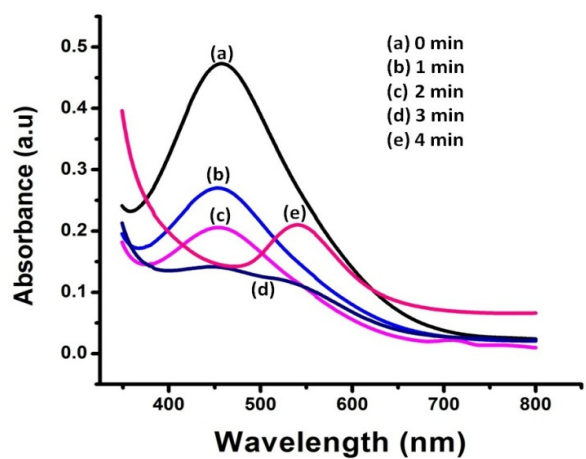

Fig. 2. Time dependent UV-Vis spectrum showing the decreasing intensity of 450 nm peak of {Mo$_{132}$} with the gradual appearance of the SPR band of AuNP-SOMs at 542 nm.

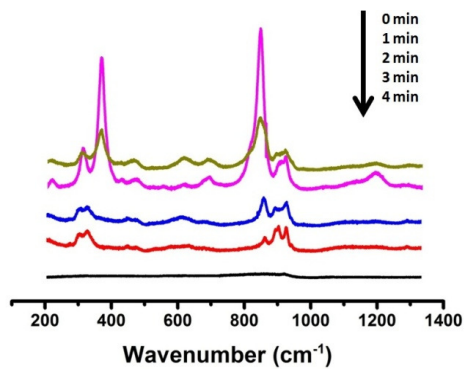

Fig. 3. Time dependent Raman spectrum of AuNP-SOM.

### 3.3. *On Controlling the morphology of AuNP-SOMs.*

To understand the effect of {$Mo_{132}$} and $HAuCl_4$ concentrations in solution on the morphology of AuNP-SOMs a series of concentration dependent experiments were carried out systematically. We hypothesized that it might be possible to manipulate the overall formation kinetics of gold nanostructures by merely changing some operational parameters of the system, thus providing us with a "control knob" to tune the size and shape of the nanostructures. To test this hypothesis we vary the molar ratio (Ψ) defined as an excess parameter with the following formulation: Ψ = [$HAuCl_4$] / [{$Mo_{132}$}]. The Ψ values were systematically varied from 20 to 40 by keeping the concentration of {$Mo_{132}$} fixed at 9 μmoles. After mixing the two stock solutions to obtain the desired Ψ value, the reaction proceeded rapidly to completion, with the solution changing color from brown to pink/violet/black within 10 to 15 minutes of stirring at room temperature. The exact color of the final colloidal gold solution depends on Ψ value. For further characterization of the so formed AuNP-SOM nanostructures, the solid was centrifuged out of the reaction mixture, washed with copious amounts of distilled water and then redispersed in distilled water by sonication. Fig 4 shows Surface Plasmon Resonance (SPR) spectra of the synthesized Au nanostructures corresponding to Ψ values of 20 and 35 respectively. For Ψ value of 35, a broad SPR band was observed centered at 520 nm (Fig 4a). This is indicative of a typical dipole resonance associated with spherical or quasi spherical Au nanoparticles.[46,47] Fig 5 is the TEM image corresponding to the SPR spectrum for Ψ = 35. The AuNP-SOMs are spherical in shape with an average diameter of 10 nm, which clearly corroborates the SPR spectrum. It was observed that the AuNP-SOMs are crystalline and their crystallinity and even gold lattices are visible from HR-TEM images and SAED-mode of HR-TEM. (Fig. 5 inset).

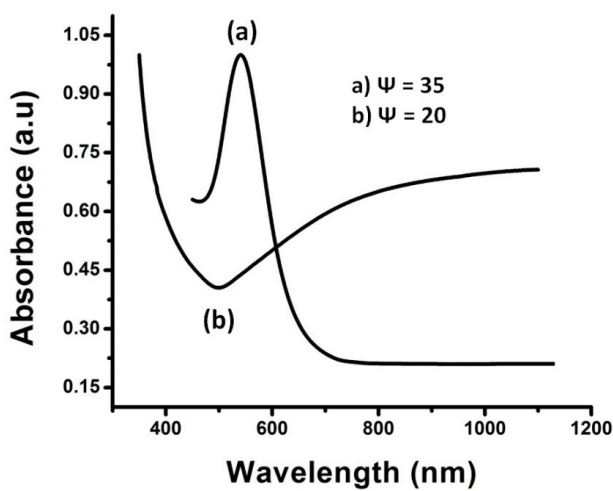

Fig. 4. SPR spectra of AuNP-SOM with different values of excess parameter Ψ. (a) Ψ=35; (b) Ψ=20.

Increasing Ψ value from 35 to 40 has no effect on the SPR band and size or shape of the nanoparticles. For Ψ greater than 40 no nanoparticles are detected in solution. New and interesting observations are made for Ψ < 30. For example, another significantly interesting effect was observed for Ψ=20. The color of the colloidal solution turns bluish-black and the corresponding UV-Visible spectrum shows a very large and wide SPR band extending from the near-IR into the whole visible range (Fig. 4b). Note that the absorbance keeps increasing from roughly 500 nm to 1200 nm, which is the upper limit of the spectral domain that could be safely explored with our UV–vis spectrometer in water. Such a broad peak at higher wavelengths corresponds to in-plane plasmon resonance band of anisotropic AuNP-SOM nanostructures.[48-50]

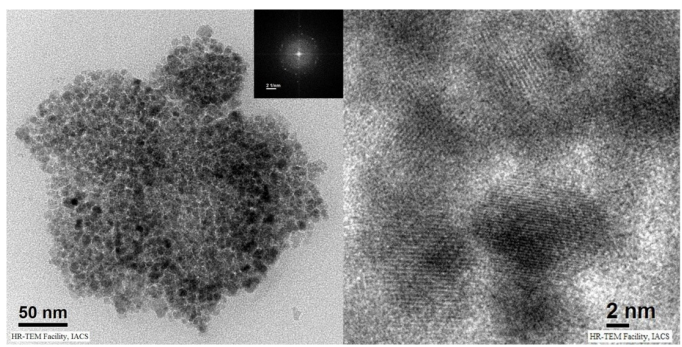

Fig. 5. TEM images of AuNP-SOM (left) and the SAED pattern (inset), lattice spacing in nanoparticles (right) for Ψ=35

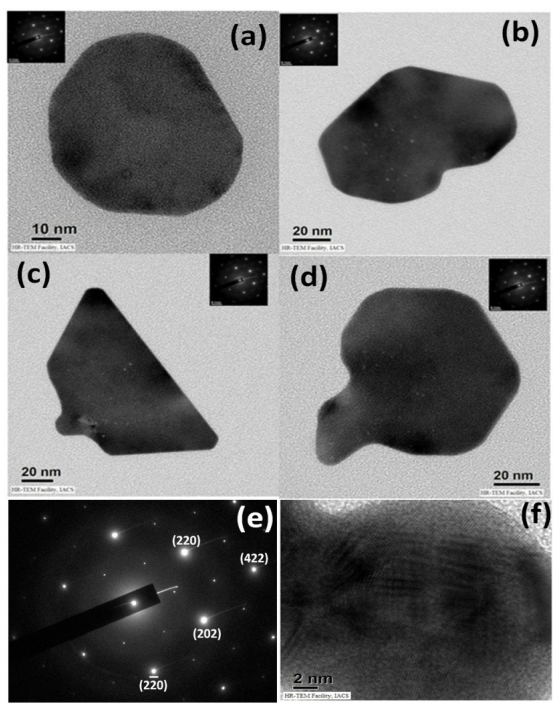

Fig 6. (a - d)TEM images of AuNP-SOM; (e) the SAED pattern of the hexagonal plate; (f)lattice spacing in nanoparticles for Ψ=20.

We now confirm the anisotropic structures from HR-TEM analysis. TEM analysis for Ψ = 20 (Fig. 6a to 6d) shows AuNP-SOM nanostructures with anisotropy, such as triangular and hexagonal plates and is in complete agreement with the corresponding SPR spectrum. Moreover, these anisotropic AuNP-SOMs are crystalline and their crystallinity and gold lattices are visible from HR-TEM images (Fig. 6f) and SAED-mode of HR-TEM (Fig. 6e). The diffraction spots of the SAED pattern can be indexed as the diffractions of a [111] zone axis, indicating that the surface of the hexagonal nanoplate is a (111) lattice plane.[51] Additionally, elemental analysis of the as-synthesized AuNP-SOMs was also performed using Energy Dispersive X-ray Spectroscopy (Fig. 7)

### 3.4. *On Synthesis and characterization of monodisperse polystyrene microspheres using AuNP-SOMs as photocatalyst*

We now test the activity of the AuNP-SOMs as photo-catalyst, since the photo-catalytic activity of several POMs is known. Recently, our group demonstrated the use of quarternerized imidazolium [α-$PW_{12}O_{40}^{3-}$] salt as a versatile recoverable photo-polymerization catalyst.[52] So, we ask, can the AuNP-SOMs, so synthesized, be also employed to act as a photo-catalytic precursors for polymerizing simple vinyl monomers under UV irradiation? We indeed observe that the as prepared AuNP-SOMs effectively catalyze the polymerization of styrene under UV light

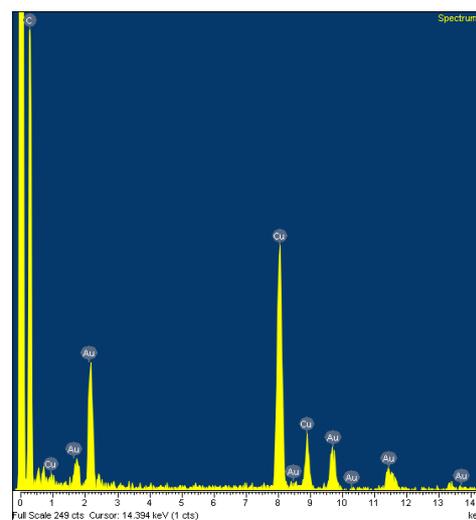

Figure 7. EDX spectrum of as prepared AuNP-SOMs.

irradiation in the absence of any external co-initiator and under open atmosphere conditions to form monodisperse polystyrene microspheres. This shows that our design strategy for synthesizing photoactive

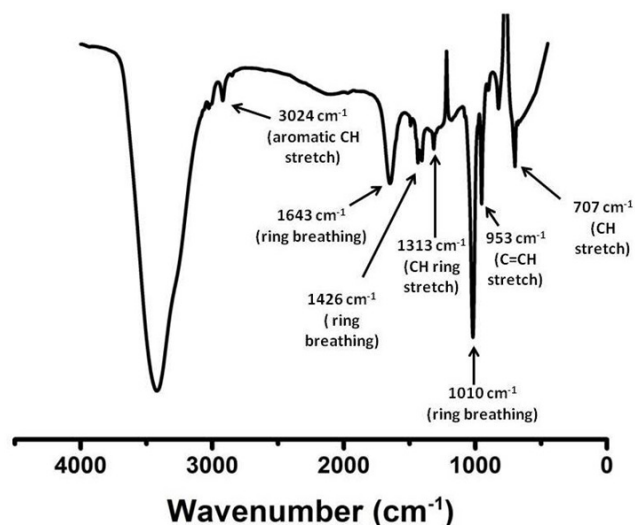

Fig. 8: ATR-FTIR spectrum of polystyrene latex

SOMs is very effective and this particular approach gives us a highly efficient photo-polymerization catalyst. Upon addition of styrene to the DMSO dispersion of AuNP-SOM, followed by irradiation with UV light, the solution starts turning turbid and eventually turns milky white. This is attributed to the generation of polystyrene microspheres in dispersion and scattering of light by these colloidal polystyrene particles. The as prepared polystyrene latex microspheres were further characterized by attenuated total reflectance fourier transform infrared red spectroscopy (ATR-FTIR). Fig. 8 shows one such typical spectrum. Some of the characteristic vibrational frequencies (in $cm^{-1}$) were assigned as follows: 1010 (C=C ring stretch); 953 (C=CH stretch); 840 (CH stretch); 1313 (CH ring stretch); 1426, 1643 (ring breathing mode); 2920, 3024, 707 (aromatic CH stretch).

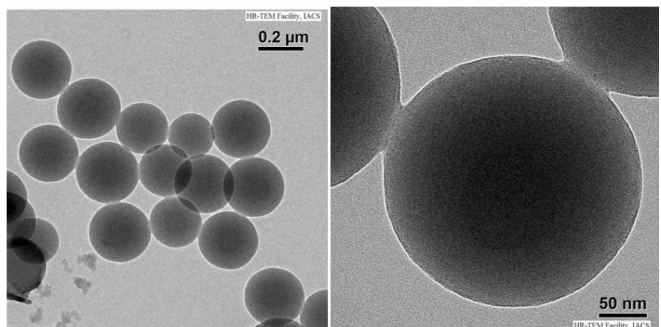

Fig.9. TEM images of as synthesized polystyrene microspheres with lower (left) and higher (right) resolutions

Fig. 9 shows the TEM images of as prepared polystyrene microspheres. All the particles were found to be uniformly spherical in shape with a diameter of approximately 200 nm. Dynamic light scattering was used to determine the mean particle size in solution. The mean hydrodynamic radius of the colloidal polystyrene microspheres was determined to be ca. 180 nm which is consistent with that of the TEM dimensions. Fig. 10 shows a typical size distribution profile of polymeric latexes with DLS.

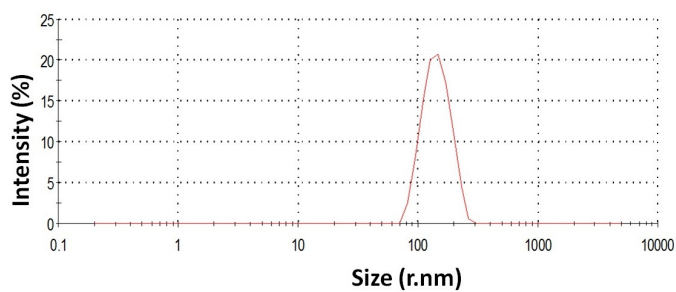

Fig. 10: Particle size distribution profile of polystyrene latex with DLS

### 3.5. *On controlling the size of polystyrene microspheres*

We now ask, is it possible to control the size of the polystyrene microspheres? To do so we test the effect of styrene concentration on the size of the microspheres. Figure 11 shows the optical microscope images of polystyrene microspheres synthesized with different initial concentrations of styrene and a constant AuNP-SOM loading of 5 wt%. The amount of styrene was varied as 200μl, 500μl, 700μl and 1ml which led to PS microspheres with varying sizes. As shown in Figure 11a, for a styrene concentration of 200μl with 5% AuNP-SOM loading the microspheres have a comparatively uniform size distribution. With an increase in styrene concentration to 500μl the PS microspheres exhibit larger sizes as shown in Figure 11b. Further increase in styrene concentration leads to further increase in particle size and still broader particle distribution coupled with inter-particle aggregation (Figure 11c, 11d). These results indicate that as the monomer concentration is increased, the AuNP-SOM photocatalytic system tends to generate larger spherical particles. We also show the corresponding hydrodynamic radii at various concentrations of styrene for a fixed loading of the catalyst in Table 1. Thus, by merely varying the initial concentration of styrene we were able to synthesize highly stable polystyrene microsphere colloids with controlled sizes under UV irradiation with AuNP-SOMs as the photocatalyst and in the absence of any external co-initiator or inert atmosphere.

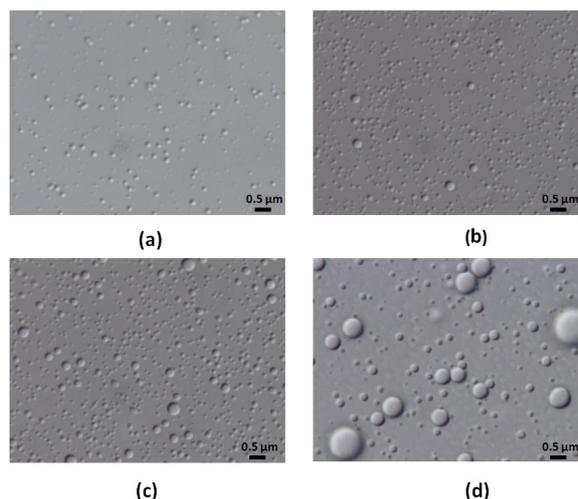

Fig. 11. PS microspheres with different initial concentrations of styrene. (a) 200 μl ; (b) 500 μl ; (c) 700 μl ; (d) 1000 μl.

Table 1. Variation of mean particle size with styrene concenration

| Amount of styrene (μl) | Mean particle Size (nm) |
|---|---|
| 200 | 180 |
| 500 | 370 |
| 700 | 610 |
| 1000 | 830 |

### 3.6. *Control Experiments*

Control experiments were performed to establish the role of AuNP-SOM as a photo-polymerization catalyst. The polymerization does not occur in the absence of AuNP-SOM or in the absence of UV irradiation. This clearly proves the role of AuNP-SOM as an efficient photo-polymerization catalyst. In order to further understand the role of oxometalate moieties, which cap the gold core, in the photo-catalytic activity of AuNP-SOM, the photo-polymerization reactions were performed using citrate stabilized gold nanoparticles instead of AuNP-SOM. However, no polystyrene was detected in the solution even after prolonged irradiation. Putting together all these evidences, following pathway is proposed (figure 12) for the formation of polystyrene microspheres by UV assisted photo-polymerization of styrene monomer using AuNP-SOM as a photocatalyst.

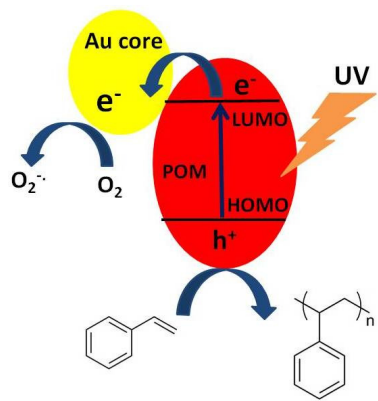

Fig. 12. Schematics of polymerization reaction by Au-NP SOM.

In short, it can be said that the oxometalates (OM) of AuNP-SOMs are known to absorb strongly in the UV region. Electron hole pairs are generated in the OM part of AuNP-SOM on UV excitation. The photo-generated electrons migrate from the LUMO of the oxometalate to the gold core of AuNP-SOM. The electron migration produces a more efficient charge separation due to the presence of surface barrier between OM and Au core and reduces the efficiency of the hole-electron annihilation rate which is an undesired process. Furthermore, because the conductivity of Au core is much higher than that of OM, the transport rate of the electrons in Au core is fast, which further reduces the recombination rate of the photo-generated electrons and holes and improves the catalytic activity of the catalyst. The photo-generated electrons are transferred to $O_2$ in the solution generating $O_2^-$. Meanwhile, the photo-generated holes oxidize and polymerize the adsorbed styrene monomer to polystyrene. The different reaction steps of this process are illustrated in figure 12.

## 4. Conclusion

Using a unique Keplerate type cluster {$Mo_{132}$} we have shown that it is possible to reduce $AuCl_4^-$ to form AuNP-SOMs dispersion which is stabilized by oxometalate corona. We further exploit the photo-catalytic activity of the so-synthesized AuNP-SOMs to synthesize polystyrene latex microspheres without any externally added initiators or inert conditions. Controlling the concentration of the monomer (styrene) we are able to control the size of the latexes. This work could to lead to a novel, facile and green route for the synthesis of latex microspheres for manifold applications in materials science.

### Acknowledgments

The authors thank DST-Fast-track fellowship, BRNS DAE grant, start-up grant from IISER-K and would like to acknowledge Ms. Deng Yinyin of CIT, China, for her kind help with optical microscope imaging of polystyrene microspheres.